\documentclass[11pt,oneside,doublespace]{amsart}
\pdfoutput=1
\usepackage[numbers,merge]{natbib}
\usepackage{geometry}                
\usepackage{graphicx}
\usepackage{amssymb}
\usepackage{amsaddr}
\usepackage{microtype}
\usepackage{caption}

\usepackage[T1]{fontenc}

\newcommand{\beginsupplement}{%
        \setcounter{table}{0}
        \renewcommand{\thetable}{S\arabic{table}}%
        \setcounter{figure}{0}
        \renewcommand{\thefigure}{S\arabic{figure}}
        \setcounter{equation}{0}
        \renewcommand{\theequation}{S\arabic{equation}}%
     }

\title[]{Experimentally Probing \textit{Hydrophobic} Water at the Gold Electrode/Aqueous Interface}
\author[Tong et al.]{Yujin Tong$^{*}$, Fran\c{c}ois Lapointe, Martin Th\"amer, Martin Wolf, and R.\ Kramer Campen$^{\dag}$}
\email[]{{*}tong@fhi-berlin.mpg.de}
\email[]{{\dag}campen@fhi-berlin.mpg.de}
\address{Fritz Haber Institute of the Max Planck Society, 4-6 Faradayweg, Berlin, Germany}                                      

\begin{document}
\maketitle

\begin{abstract}
Quantitative description of reaction mechanisms in aqueous phase electrochemistry requires experimental characterization of local water structure at the electrode/aqueous interface and its evolution with changing potential. Gaining such insight experimentally under electrochemical conditions is a formidable task. Here we characterize the potential dependent structure of a sub-population of interfacial water that have one OH group pointing towards a gold working electrode using interface specific vibrational spectroscopy in a thin film electrochemical cell. Such \textit{free-OH} groups are the molecular level observable of an extended hydrophobic interface. This \textit{free-OH} interacts only weakly with the Au surface at all potentials, has an orientational distribution that narrows approaching the potential of zero charge and disappears on the oxidation of the gold electrode.
\end{abstract}

\maketitle

\vspace{0.2cm}

Water controls aqueous phase electrochemistry: it forms the electric double layer, determines solute/electrode association, and can accept and donate electrons. These processes depend strongly on interfacial water's distance from, and orientation with respect to, the electrode surface \citep{mar64,*dog75}. In liquid water such \emph{\textbf{local}} structure is largely determined by water's hydrogen bond network. Characterizing this network at electrode surfaces, and its change with potential, is thus a prerequisite to quantitative insight into virtually all aqueous electrochemistry \cite{vel14,gro14,kol16b,*mai15,*car12}.

Electrostatics suggests that water at electrode surfaces is on average orientated with its hydrogens pointing towards negatively and away from positively charged electrodes\cite{boc00}. Simulation studies suggest that while this insight is true, water structure at electrode surfaces is substantially more intricate \cite{gro14,sch96,izv01,*ota08,*sch09}: \textit{e.g.}\ OH groups point both towards and away from metals at all potentials and interfacial water density is higher at positively than negatively charged electrodes.

Gaining similar experimental insight is challenging. Both x-ray scattering (XRS) and surface enhanced infrared (SEIRAS) studies of water's bending mode confirm the potential dependent orientation picture suggested by electrostatics, while XRS studies also find that interfacial water density increases going from negative to positive potentials \cite{sch96,ton94,ata96}. However both approaches only provide structural information averaged in the plane of the interface: XRS because of the high, non-resonant photon energies it typically employs and SEIRAS of water's bend because of the relative insensitivity of the bend to local structure.

In principle, water's OH stretch spectral response reports on local H-bonding structure \cite{gei13}. However, attempts to gain such insight in a SEIRAS scheme, because of the delicate referencing such measurements require, and have been only able to show interfacial water's \emph{average} H-bond strength increased going from negative to positive potentials\cite{ata96}. Recently Salmeron and coworkers combined x-ray absorption(XRA) spectroscopy and simulation to address water's potential dependent H-bond structure at a gold electrode\cite{vel14}. While simulation clearly illustrated multiple types of H-bonded interfacial water, and their change in relative densities with potential, XRA measurements were sensitive to only one type of interfacial water (< 1/3 of water present); that is, water molecules donating one hydrogen bond to other water molecules whose non-H-bonded OH group lies parallel to the Au surface.

In this study, we characterize water structure at the gold electrode (chosen because it is relatively inert \cite{mic06}) using the spectral response of the OH stretch of interfacial water. We access this observable using interface-specific vibrational spectroscopy (vibrational sum frequency (VSF) spectroscopy) and characterize a population of interfacial water molecules with one OH pointing towards the Au surface and interacting with it weakly. Simulation studies predict more than 1/3 of interfacial waters are of this type\cite{vel14}. 

To record a VSF spectrum we spatially and temporally overlap the output of pulsed mid-infrared (IR) and 800 nm (VIS) lasers at the Au/solution interface in a thin-film spectroelectrochemical cell, and monitor the output at the sum of the frequencies (VSF) of the two incident fields. The VSF signal is interface specific by its symmetry selection rules and a spectroscopy because the emitted VSF intensity increases by several orders of magnitudes when the frequency of one of the incident fields, in this case the IR, is in resonance with a vibrational mode at the interface\cite{she89}. All beams copropogate in a plane normal to the electrode surface and reach the interface through the thin liquid film (see Supporting Information for detailed description of our laser system, sample cell geometry and control experiments). 

\begin{figure}
	\includegraphics[width=0.55\textwidth]{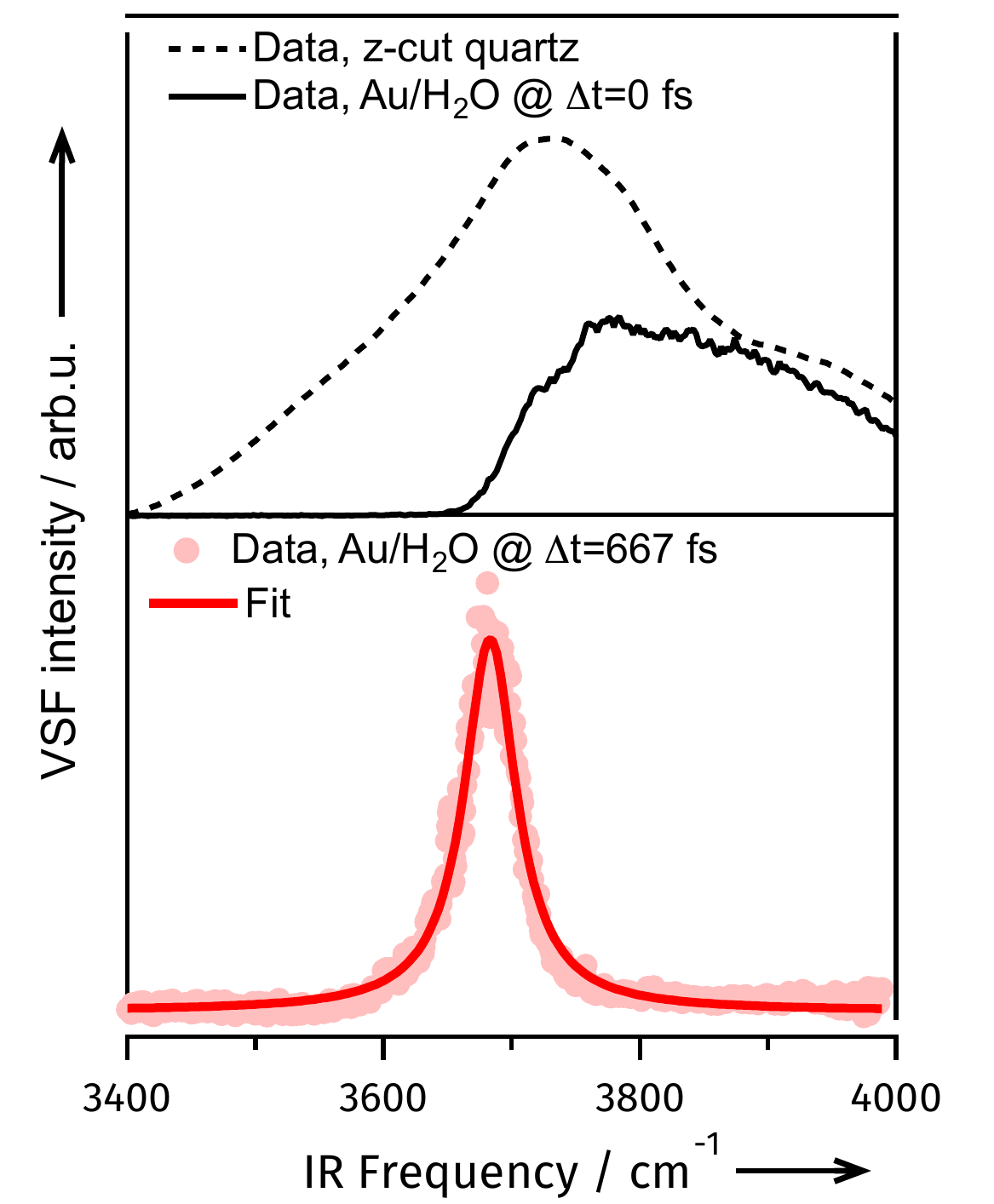}
	\caption{(top) Dashed line: profile of the incident IR laser as measured from z-cut quartz; Solid line: VSF spectrum from the Au/solution interface at 0 delay between the VIS and IR lasers. (bottom) Pink dots: VSF spectrum from the gold/H$_2$O interface with the 800 nm beam delayed 667 fs with respect to the IR. Solid red curve: fit of the data with a single Lorentzian resonance. All measurements were done in the \textit{ppp} (\textit{p}-\textsc{vsf}, \textit{p}-\textsc{vis}, \textit{p}-\textsc{ir}) polarization combination.}\label{ocp}
\end{figure}

The VSF spectrum from the gold/solution interface at open circuit potential, plotted vs. frequency of the incident IR field, is shown in Figure \ref{ocp}(top). Absent significant absorption of the VIS or IR beams before the sample or the emitted VSF after the sample but before detection, much prior work suggests that the measured response should be dominated by the non-resonant response of the gold surface \cite{bac12}. However, because all light goes to and from the interface through solution in our sample cell we find, as comparison to the non-resonant quartz signal in an air-filled cell makes clear, that essentially all incident infrared intensity between 3400 and 3600 cm$^{\text{-1}}$ is absorbed by the OH stretch of bulk water (see Figure \ref{ocp}(top)). Nevertheless, comparison of the two spectra in Figure \ref{ocp}(top) clearly shows that significant IR intensity passes through the thin water layer at energies above 3600 cm$^{-1}$ and show a suggestive shoulder at 3700 cm$^{-1}$. To say more about this feature we suppress the nonresonant Au signal by temporally delaying the VIS up-conversion pulse relative to the IR\cite{lag07}. In principle this delay suppresses the spectral response of the rapidly dephasing non-resonant signal and preserves that of more slowly dephasing resonances. Indeed, delaying the 800 nm upconversion beam by 667 fs with respect to the maximum IR intensity produces a spectrum with a narrow peak centered at $\approx$ 3680  cm$^{-1}$, see Figure \ref{ocp}(bottom).

While effective at suppressing non-resonant contributions, this approach has been shown to introduce system dependent spectral distortions \cite{sti10}. We performed a series of control experiments to evaluate whether such distortions could be responsible for our observed signal (see Supporting Information for details and relevant data). These experiments clarify that the narrow spectral feature we observe is a slowly-dephasing resonance at the Au electrode/water interface.

\begin{figure}[h!]
	\includegraphics[width=0.58\textwidth]{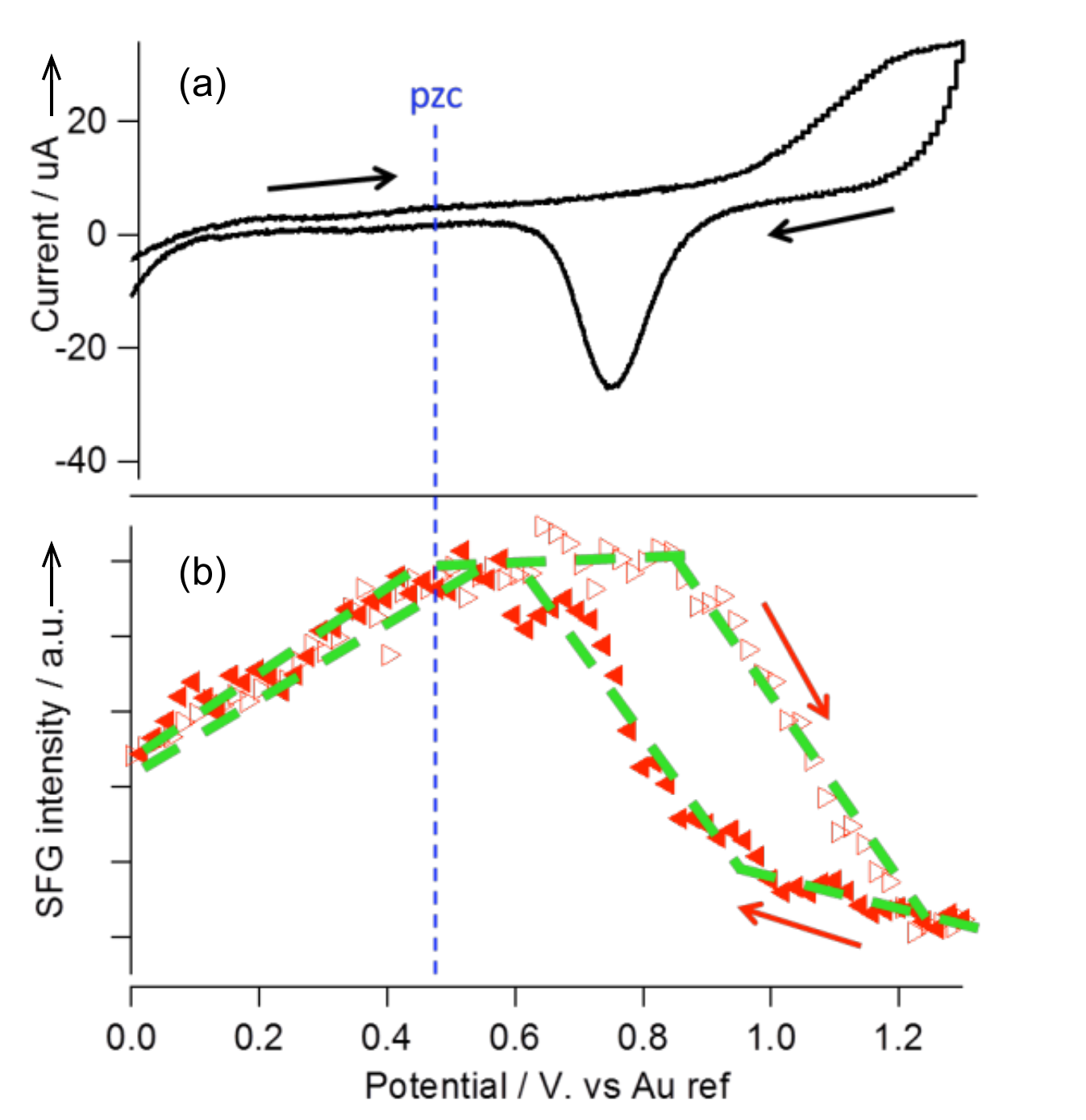}
	\caption{a) Cyclic voltammogram of a gold electrode in 1 M HClO$_4$ aqueous solution with a scan rate of 5 mV/s and flowing electrolyte. An evaporated Au thin film was used as an internal reference. b) The potential-dependent VSF intensity of the free-OH peak during the same scan. Arrows are direction of the sweep. Dashed blue line is the measured pzc.}\label{pot_dep}
\end{figure}

A fit of the H$_2$O signal at open circuit potential following previously described procedures \cite{lam05} yields a center frequency of 3680 cm$^{-1}$ and full width half maximum of 44 cm$^{-1}$. Similar OH stretch features have been observed previously at a variety of water/hydrophobic interfaces and been assigned to {\it free}-OH groups: OH groups originating from water molecules that straddle the interface with one pointing away from bulk liquid water and the other toward with only the latter donating an H-bond \citep{ton13,*du94,*sca01}. Because the two OH stretch vibrations on this type of water differ in frequency by 150 cm$^{-1}$ they are decoupled and the transition dipole of the free-OH stretch response is 53$^{\circ}$ from the water molecule's permanent dipole and along the OH covalent bond. These \textit{free}-OH groups are a known to be a microscopic (\textit{i.e.}\ molecular-level) characteristic of extended hydrophobic surfaces \cite{cha05}. Because we are interested in potentials well below water dissociation on Au, and because there is no evidence in prior simulation studies for interfacial water molecules whose OH groups point towards the liquid but are non-H-bonded \cite{vel14}, we assign this feature to a similar under H-bonded water with one OH group pointing away from the liquid and only weakly interacting with gold. Viewed from a macroscopic perspective, \textit{i.e.}\ contact angle measurements, gold becomes relatively hydrophobic with even small amounts of adsorbed carbon \cite{smi80} (although interpreting potential-dependent contact angles is challenging \cite{kan02}). Because we observe no evidence of adsorbed CH moieties in our cell (see Supporting Information for data) and no oxidation current associated with adsorbed carbon in the in-situ cyclic voltammagram (CV) (see Figure \ref{pot_dep}a) the result in Figure \ref{ocp}(bottom) clearly suggests that, at least at open circuit potential, from a microscopic perspective the gold surface is hydrophobic. Connecting this molecular-level indicator of the Au surface hydrophobicity with macroscopic measurements is an object of current research in our group.

Figure \ref{pot_dep}(a) shows the in-situ CV of the gold electrode. In agreement with previous reports, in the positive-going scan the CV is rather flat until an oxidation peak between 1.0-1.25 V (vs. Au pseudo reference); in the negative-going scan, a reduction feature appears centered at 0.75 V. The former is known to be due to Au oxidation and the latter to reduction of the Au oxide \cite{ata96,bur97}. In independent amperometery measurements (see Supporting Information for data) we find that the potential of zero charge (pzc) under the conditions of the CV was 0.5 V, consistent with prior efforts \cite{ham96b,*ham95b,*efr73}. The integrated intensity of the narrow VSF feature, collected simultaneously with the CV, is shown in Figure \ref{pot_dep}(b).  In the positive sweep direction we find an increase of the VSF intensity from 0 to 0.5 V (\textit{i.e.}\ from negative potentials to the pzc), a plateau from 0.5-0.8 V, and a rapid decrease with the onset of oxidation at still more positive potentials. In the negative-going scan the signal first weakly (1.2-0.9 V) and then rapidly (0.9-0.6 V) increases with decreasing potential as the Au-oxide is dissolved. At potentials negative of the pzc (0.6-0 V), the integrated VSF intensity of both scans are equal. Because spectral shape changes minimally with potential, integrating intensities loses little physical insight (see Supporting Information for details). 

\begin{figure}[h!]
	\includegraphics[width=0.9\textwidth]{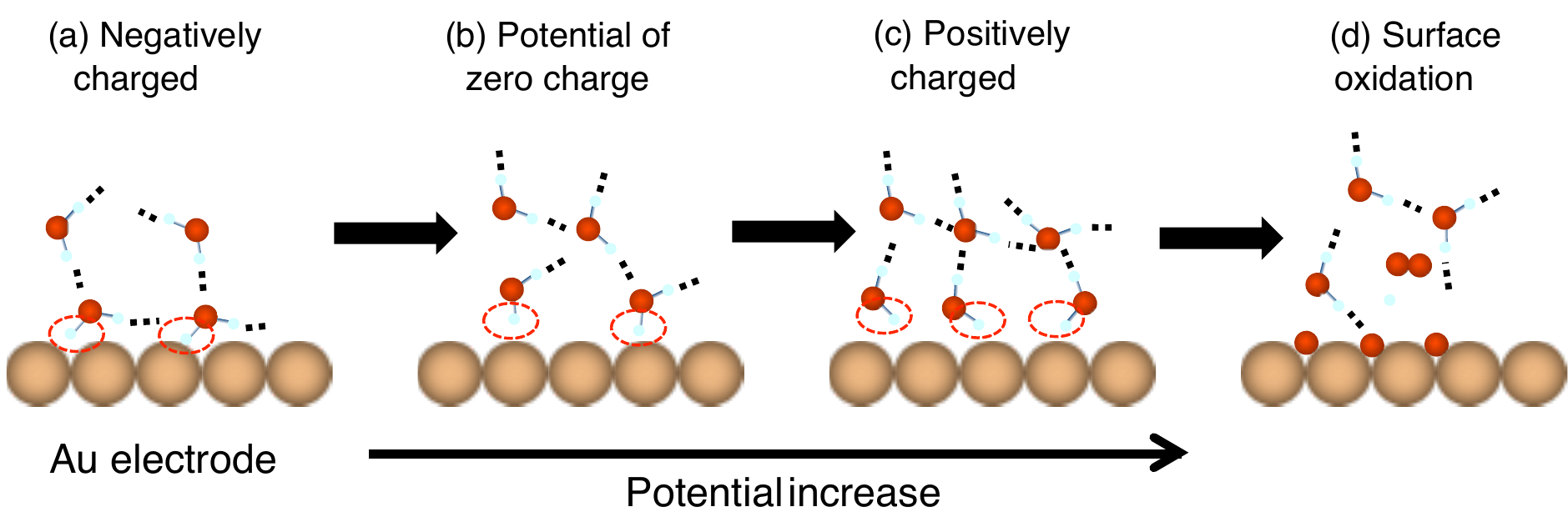}
	\caption{Potential-dependent water structure revealed by VSF measurements. Dotted lines indicate donation of H-bonds to other water molecules.}\label{cartoon}
\end{figure}

The disappearance of the free-OH spectral feature after oxidation is expected, the hydrophilic oxide surface should be a strong hydrogen bond acceptor from bulk water and the stabilizing effect of an image charge interaction is removed on oxide formation, rationalizing the potential dependence of VSF intensity in the absence of oxide formation is more challenging. Much prior work has shown that the maximum VSF signal from molecules at metal surfaces occurs under a \textit{ppp} polarization condition when the mode's transition dipole is orthogonal to the metal surface \cite{lam05}. As noted above, both theory and experiment find that at potentials positive/negative of the pzc water molecules tend to be oriented with their OH groups pointing away from/toward the metal surface due to interaction of the surface field with water's permanent dipole. Because water's permanent dipole bisects both OH groups, it is expected that at sufficiently negative potentials neither OH would be perpendicular to the electrode surface (see Figure \ref{cartoon}(a)). As potential is increased, but below the pzc, the energetic penalty for non-perpendicular water permanent dipole orientations becomes smaller, and thus we expect the probability of observing a \textit{free}-OH perpendicular to the electrode surface, and thus the VSF intensity, should grow (see Figure \ref{cartoon}(b)). This logic suggests that the VSF signal should also decrease at potentials positive of the pzc (but below oxide formation), an effect we do not observe. Note, though, that as discussed above, both experiment and theory suggest that the density of interfacial water is higher for potentials positive of the pzc than negative \cite{ton94}. Because VSF intensity is proportional to the density of interfacial OH oscillators squared, we take the plateau in intensity between 0.5 and 0.9 V to indicate a compensation of orientation and density effects.

These results are consistent with prior studies that have suggested interfacial water is on average oriented with its static dipole moment pointing towards the electrode at potentials negative of and away at potentials positive of the pzc and studies that suggest an increase in interfacial water density at potentials positive of the pzc. While important all of these insights average over all structural \textit{types} of interfacial water. The results of this study are the first experimental characterization of a population of interfacial water heretofore only predicted to exist theoretically, \textit{i.e.}\ water with one OH group perpendicular to the gold surface and not hydrogen bonded and the other donating a hydrogen bond to the liquid. Clearly there is much left to learn about potential-dependent water structure at gold electrodes, current work in our group focuses on the extension of this approach to different structural types of interfacial water using different experimental configurations \cite{liu14c} and on characterizing the role of this free-OH in a variety of aqueous photoelectrochemical processes. Nevertheless, the results presented here, in an optical experiment employing femtosecond pulses, are the first step in direct studies of the role of water in many aqueous (photo)electrochemical reactions.    

\section*{Acknowledgements}
The authors thank the Deutsche Forschungsgemeinschaft for support of this study through Collaborative Research Center 658: Elementary Processes in Molecular Switches at Surfaces.

\providecommand*{\mcitethebibliography}{\thebibliography}
\csname @ifundefined\endcsname{endmcitethebibliography}
{\let\endmcitethebibliography\endthebibliography}{}

\clearpage

\textbf{TOC Figure}
\begin{figure}[h!]
	\includegraphics[width=0.85\textwidth]{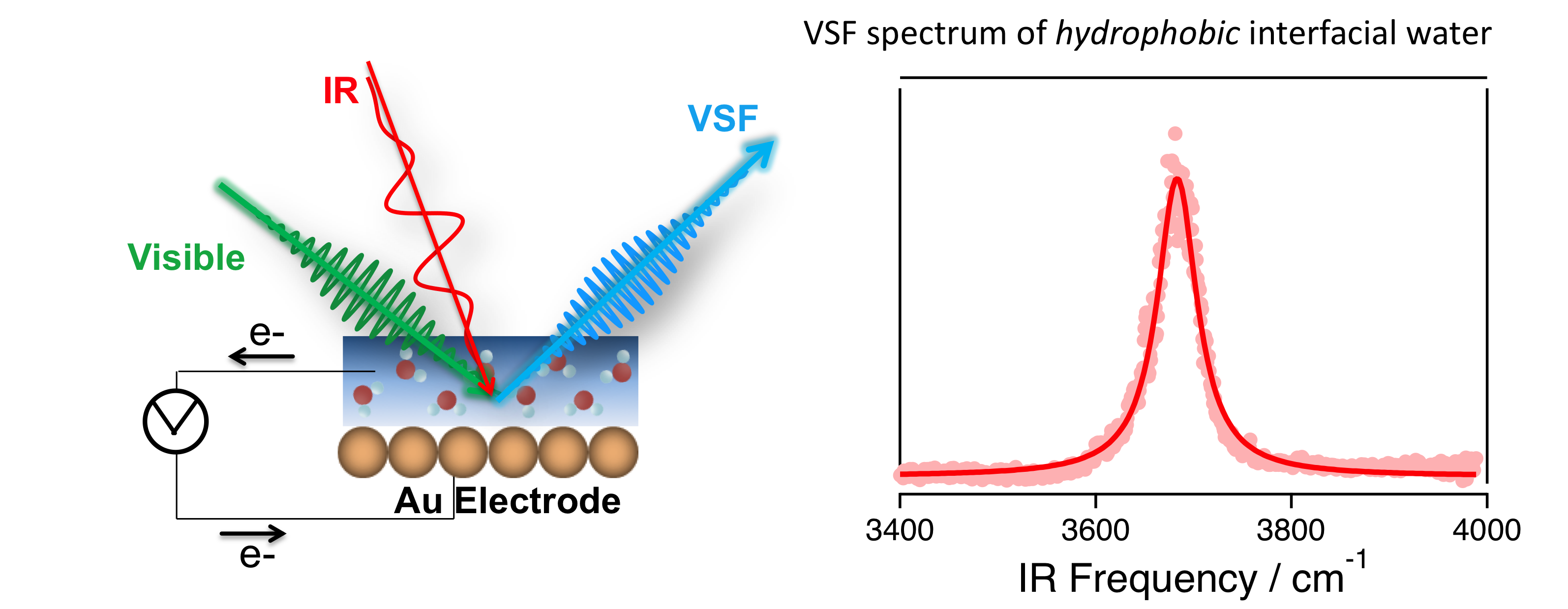}
	\caption*{A population of water molecules with one OH pointing away from bulk liquid, i.e.\ \textit{hydrophobic} water, are characterized at the gold electrode/aqueous electrolyte interface using interface-specific vibrational spectroscopy. This type of interfacial water disappears on oxidation of the gold electrode. At potentials below gold oxidation its structure is bias dependent.}\label{TOC}
\end{figure}

\clearpage

\beginsupplement

\section*{Supporting Information} 
\subsection*{Experimental Details of the VSF Measurements}
The VSF setup used in the current study has been described in our previous publications \cite{ton15,ton16}. In brief, the laser system is composed of a Ti:Sapphire oscillator (Venteon, Femtosecond Laser Technologies) and regenerative amplifier (Legend Elite Due HE and Cryo PA, Coherent). The regenerative amplifier output -- 7.5 mJ/pulse, 45 fs pulses, 1 KHz, centered at 800 nm -- is used to pump a commercial optical parametric amplifier (HE-TOPAS, Light Conversion) the signal and idler output of which are mixed in a 1 mm thick AgGaS$_2$ crystal in a non-collinear difference frequency generation scheme to produce the infrared (IR). As mentioned in the main text, the center frequency of the IR beam was tuned to 3600 cm$^{-1}$ for probing hydrophobic water species.  To generate the narrowband visible (VIS) pulse we used the residual 800 nm light from the OPA and an air spaced etalon (SLS Optics Ltd). The resulting beam has its maximum intensity at 800 nm with a bandwidth of 10 cm$^{\text{-1}}$. A band pass filter centered at 800 nm was used to filter out the high order components of the transmittance through the etalon. The energy per pulse of the IR and VIS at the sample surface was controlled using $\lambda/2$ plate, polarizer, $\lambda/2$ plate combinations and was 4.5 and 8.0 $\mu$J/pulse respectively. The two beams were directed so as to propagate in a coplanar fashion (in a plane normal to the electrode surface) and focused on the sample using lenses with focal lengths of 25 and 45 cm and incident angles of $\text{40.4}^{\circ} \pm \text{0.5}^{\circ}$ and $\text{65}\pm \text{0.5}^{\circ}$ (both with respect to the surface normal) for the IR and VIS respectively. After collimation the VSF signal was dispersed in a spectrograph (SR303i, Andor Technology) and imaged on an EMCCD camera (Newton, Andor Technology). All measurements were conducted in ambient conditions at room temperature and and the \textit{ppp} polarization condition (\textit{p}-polarized SF, \textit{p}-polarized visible, and \textit{p}-polarized IR where \textit{p} indicates polarization in the plane of incidence). The acquisition time for each spectrum was 30 s.  In all VSF measurements, we collected multiple voltammetric cycles and the spectra we record in multiple positive and negative going sweeps are identical.

To quantify the VSF spectral response, we follow prior works and describe the measured sum frequency intensity as a coherent sum of a nonresonant background and Lorentzian resonance(s) \cite{ton15,ton16}.  
	\begin{equation}
		\text{I}_{\text{vsf}} \propto \left|\left|\chi_{\text{nr}}\right|\text{e}^{i\phi} + \sum_{n}\frac{\chi_{n,ijk}}{\omega_{\text{ir}} - \omega_{n} + i\Gamma_{n}}        \right|^{2}\text{I}_{\text{vis}}\text{I}_{\text{ir}}
	\end{equation} 
where $\omega_{\text{ir}}$ is the frequency of the incident infrared, $\chi_{\text{nr}}$ and $\phi$ are the nonresonant amplitude and phase, $\chi_{n,ijk}$, $\omega_{n}$ and $2\Gamma_{n}$ are the complex amplitude, center frequency and line width of the $n^{th}$ resonance, and I$_{\text{vis}}$ and I$_{\text{ir}}$ are the 800 nm and infrared intensity respectively. By delaying the IR with respect to the VIS we suppress the amplitude of the nonresonant contribution and the resulting fits are therefore in practice dominated by the resonance.

\subsection*{Electrochemical Conditions}
\begin{figure}
	\includegraphics[width=0.75\textwidth]{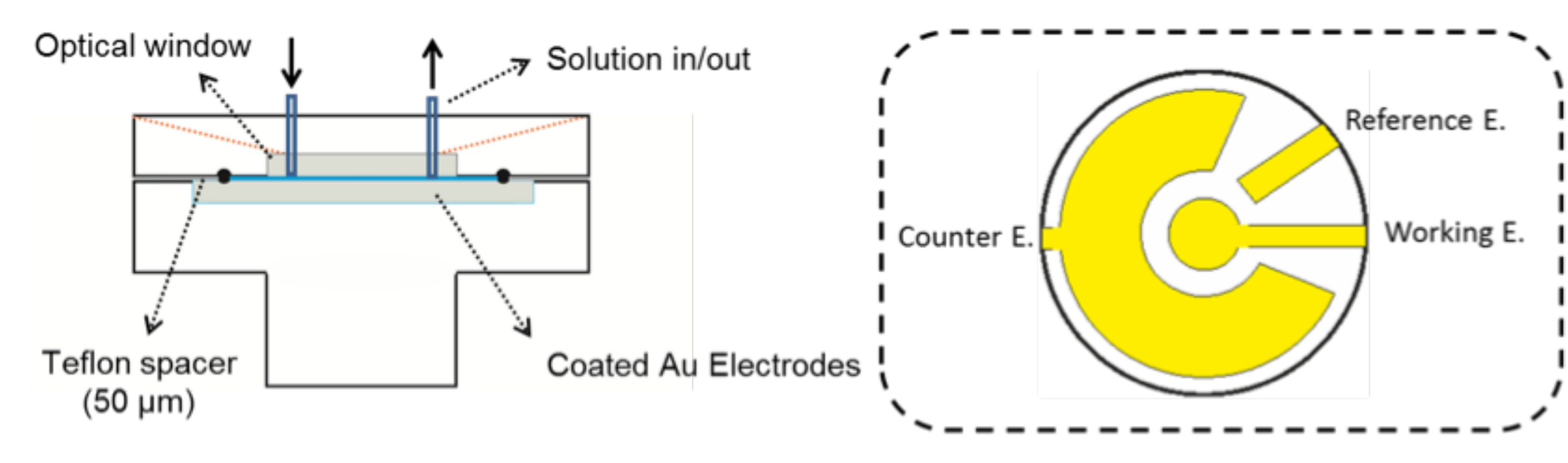}
	\caption{Thin film flow cell for spectroelectrochemical measurements.}\label{exp}
\end{figure}

We employed a home-built thin layer electrochemical cell constructed of a CaF$_2$ window and three gold electrodes vapor deposited on a z-cut qtz substrate (see Figure \ref{exp}) for all measurements. To minimize transport concerns and maximize optical signals we adopted a film thickness of 50 $\mu$m. This thickness was controlled using commercially available Teflon spacers (Goodfellow, Cambridge). A solution of 1M HClO$_4$ (Merck KGaA) in ultrapure H$_2$O (18.2 $\text{M}\Omega\cdot\text{cm}$ at 25 $^{\circ}$C, Millipore Corporation) and D$_2$O (Sigma-Aldrich) was used for measurements in O-H stretching and O-D stretching frequencies, respectively. All solutions were de-gassed with pure N$_2$ (99.999\%, Westfalen) 30 min before and throughout the measurements in a reservoir that continually supplies fresh solution to the thin layer cell. Cyclic voltammetry (CV) measurements were carried out using a VoltaLab potentiostat (PGZ301, Radiometer analytical).

A coated Au film (right side of Figure \ref{exp}) was employed as a pseudo reference during the electrochemical measurements. As the CV measured with this pseudo reference electrode is similar to that measured with conventional references, \textit{i.e.}\ standard reference electrode (SHE), but with an offset of ~0.4 V (see Figure \ref{CV})\cite{bur97}, It is, in fact, a reference in our thin film cell.
\begin{figure}
	\includegraphics[width=0.66\textwidth]{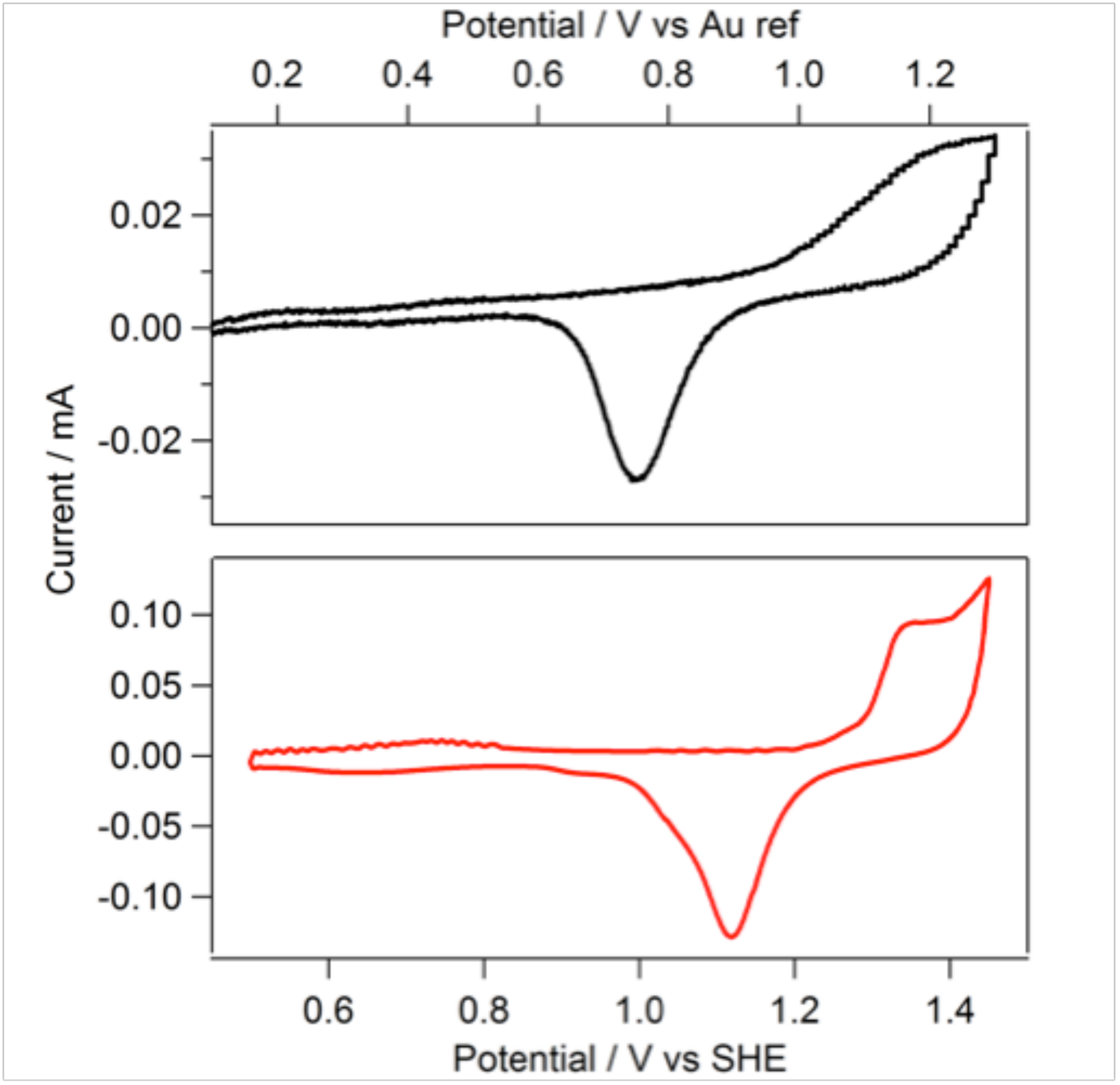}
	\caption{Cyclic voltammetry obtained in the thin film cell with and Au pseudo reference (top panel) and that in bulk solution (bottom panel) with standard hydrogen electrode reference.}\label{CV}
\end{figure}

\begin{figure}
	\includegraphics[width=0.66\textwidth]{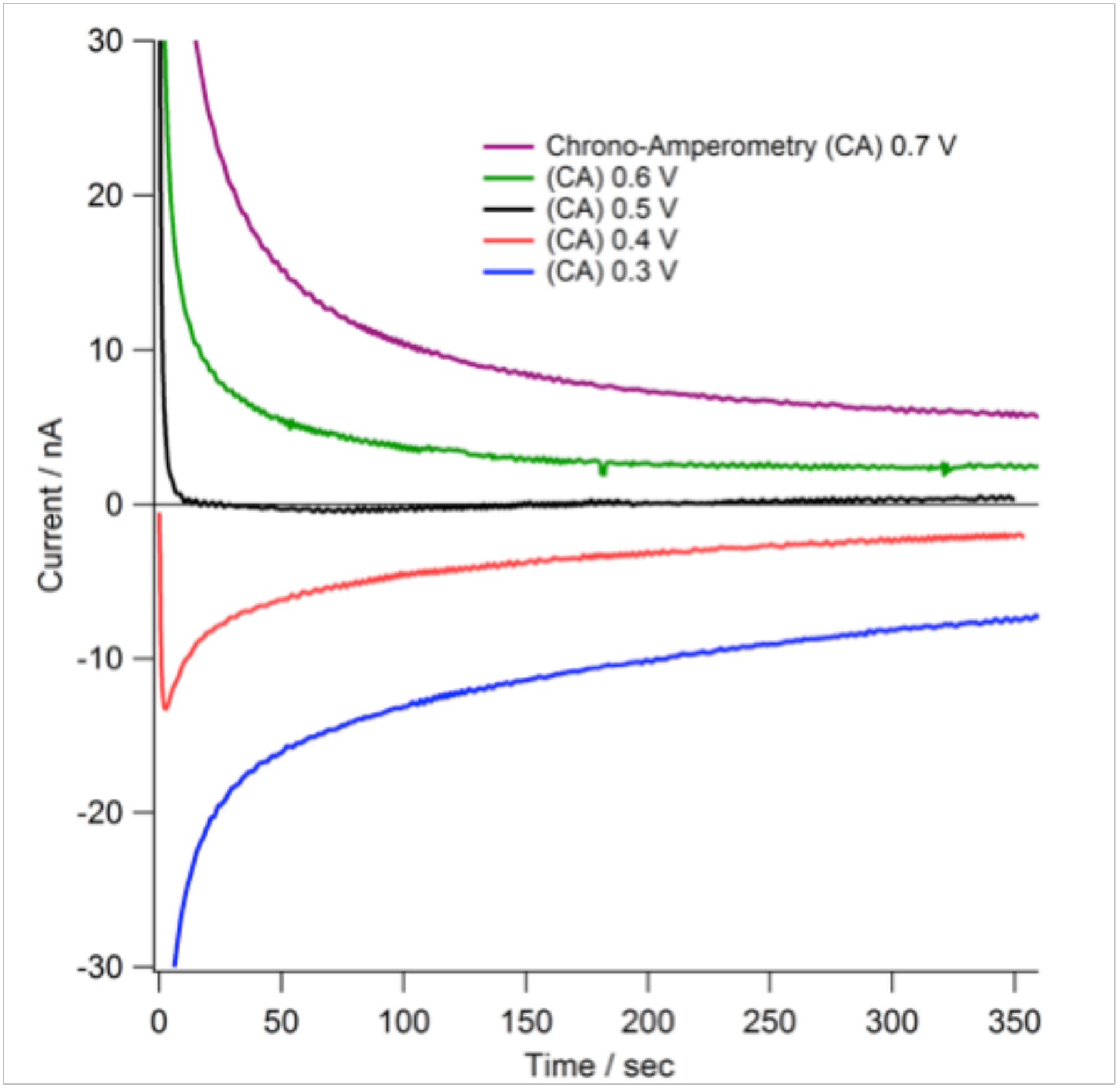}
	\caption{Chrono-amperometry measurement for determining the pzc of current electrochemical cell.}\label{amp}
\end{figure}

The potential of zero charge (pzc) of the current flow cell was determined by measuring the current in chronoamperometry at various steady potentials. Typical measured currents were in the 10$^{-9}$ A range. As shown in Figure \ref{amp}, the current is close to zero when the potential was held at 0.5 V, the approximate value of the pzc.

\subsection*{Control Experiments}
\subsubsection*{Verifying the Absence of Hydrophobic Interfacial Carbon Compounds: Contaminants}
\begin{figure}
		\begin{center}
			\includegraphics[width=0.7\textwidth]{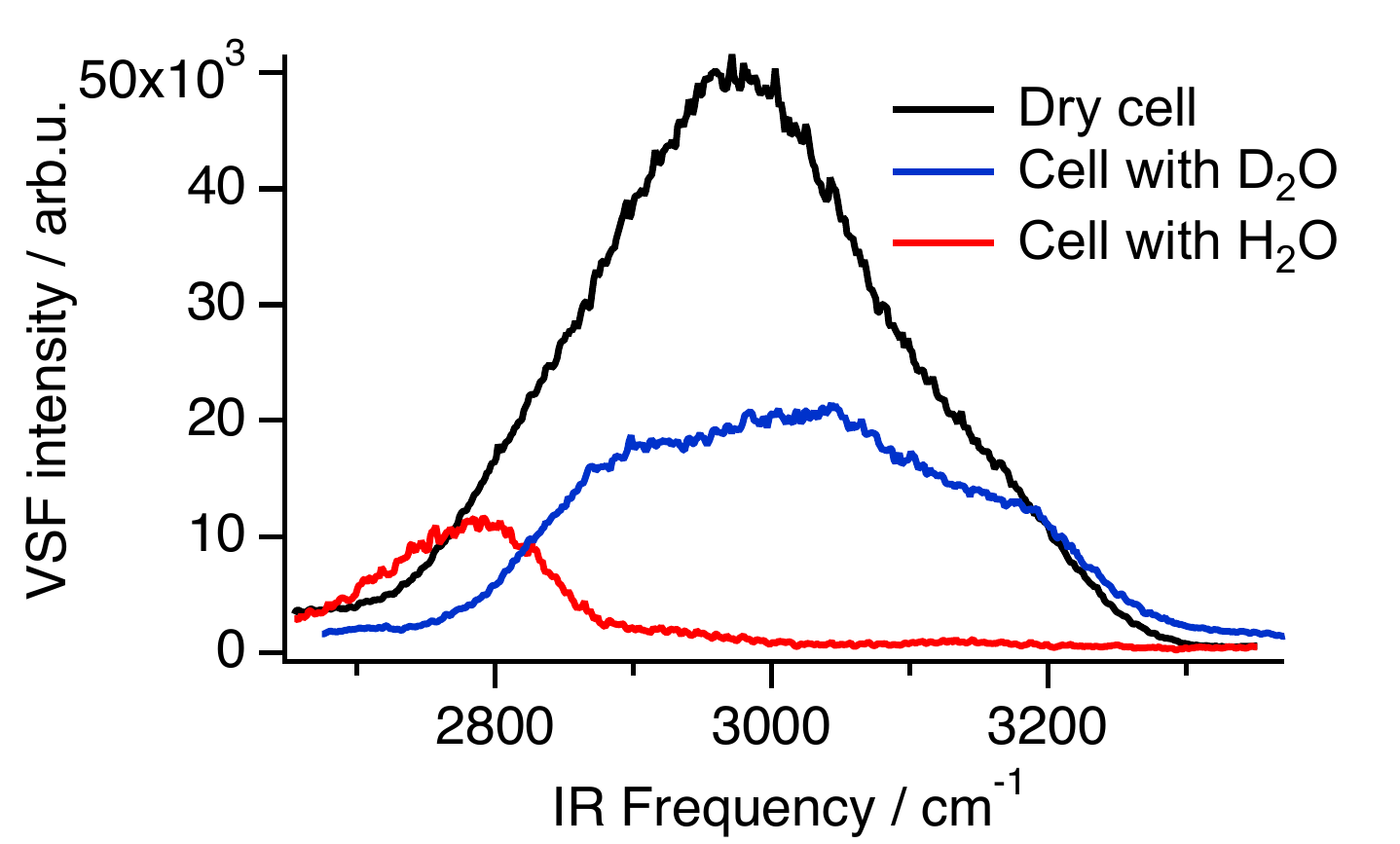}
			\caption{VSF spectra, at zero delay between IR and VIS interactions, for the dry cell, H$_{\text{2}}$O and D$_{\text{2}}$O filled cells. All spectra show a clear absence of CH stretch related spectral features.}
		\label{CH}
		\end{center}
	\end{figure}

As noted in the main text the hydrophobicity of Au, as measured via contact angle, is observed to depend critically on the presence of small amounts of organic contaminants \cite{smi80}. We tested for the presence of such adsorbed, hydrophobic, compounds in two ways: we took VSF spectra in the CH stretch region in a dry, H$_{\text{2}}$O filled, and D$_{\text{2}}$O filled cell and conducted an in-situ cyclic voltammagram while collecting VSF spectra. The control VSF spectra are shown in Figure \ref{CH} and clearly show only the nonresonant response of Au (for the H$_{\text{2}}$O filled cell there is essentially no observed VSF intensity above 3000 cm$^{\text{-1}}$ because of IR absorption by the aqueous electrolyte). As discussed above, the in-situ cyclic voltammagram we collected off of Au quantitatively agrees with prior work. The details of this voltammagram change significantly even in the presence of small amounts of adsorbed carbon. To give a sense for these changes we show below a CV collected in our cell in the same electrolyte, over the same potential range, in the presence of an additional 0.1 M formic acid (see Figure \ref{CV}). The additional oxidation features at potentials below 1.4 V are known to be the result of formate adsorption/oxidation and clearly indicate the sensitivity of this set up to even small amounts of carbon contamination \cite{cue13}.   
	
	\begin{figure}
		\begin{center}
			\includegraphics[width=0.7\textwidth]{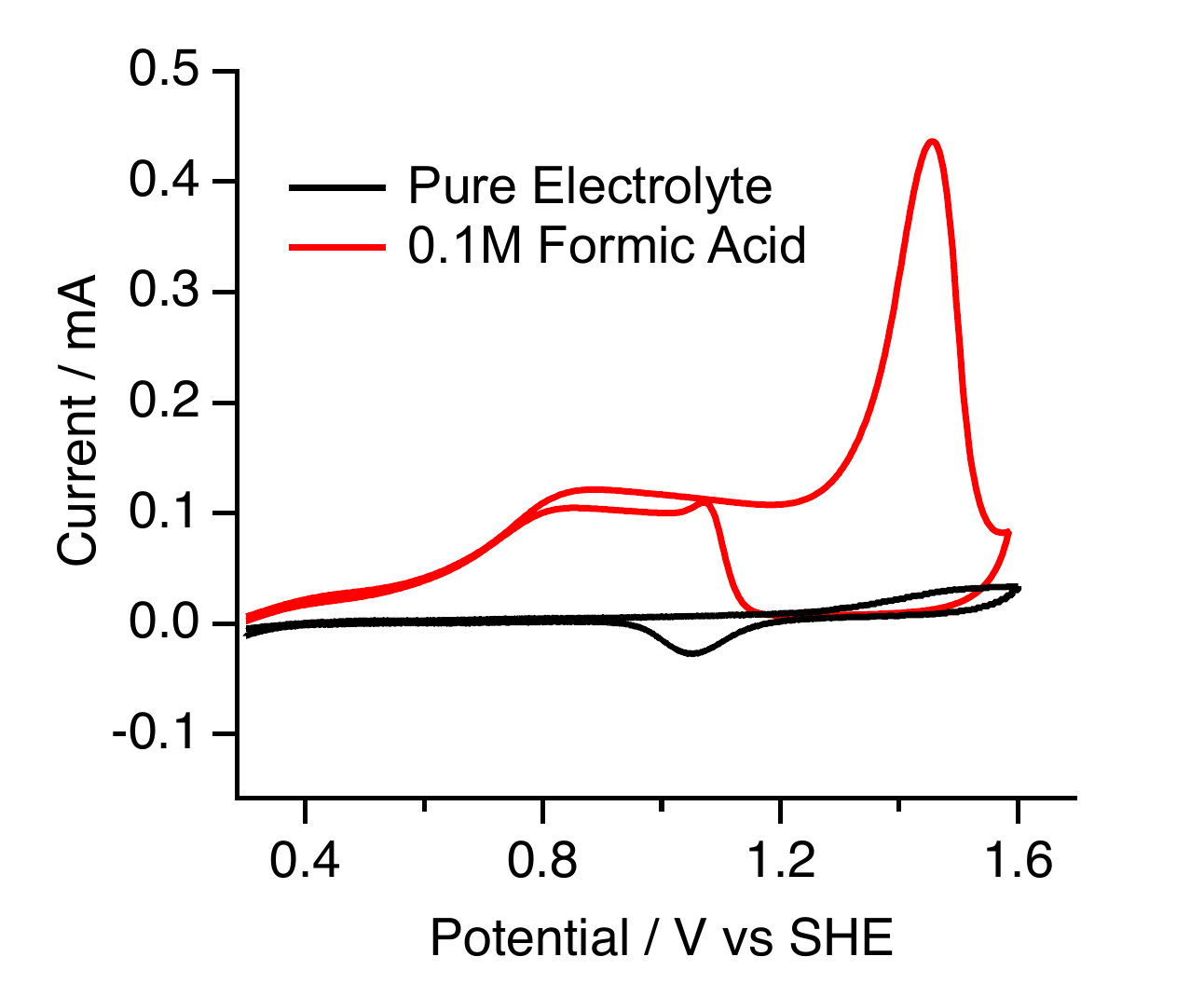}
			\caption{Cyclic voltammagram collected under the conditions of the experiment, 1 M HClO$_{\text{4}}$ electrolyte in ultrapure water, in our thin film spectroelectrochemical cell, and under the same conditions with 0.1 M formic acid.\label{CV}}
		\end{center}
	\end{figure}

\subsubsection*{Effect of VIS-IR Delay on Observed Spectral Feature}
\begin{figure}[h]
	\includegraphics[width=0.60\textwidth]{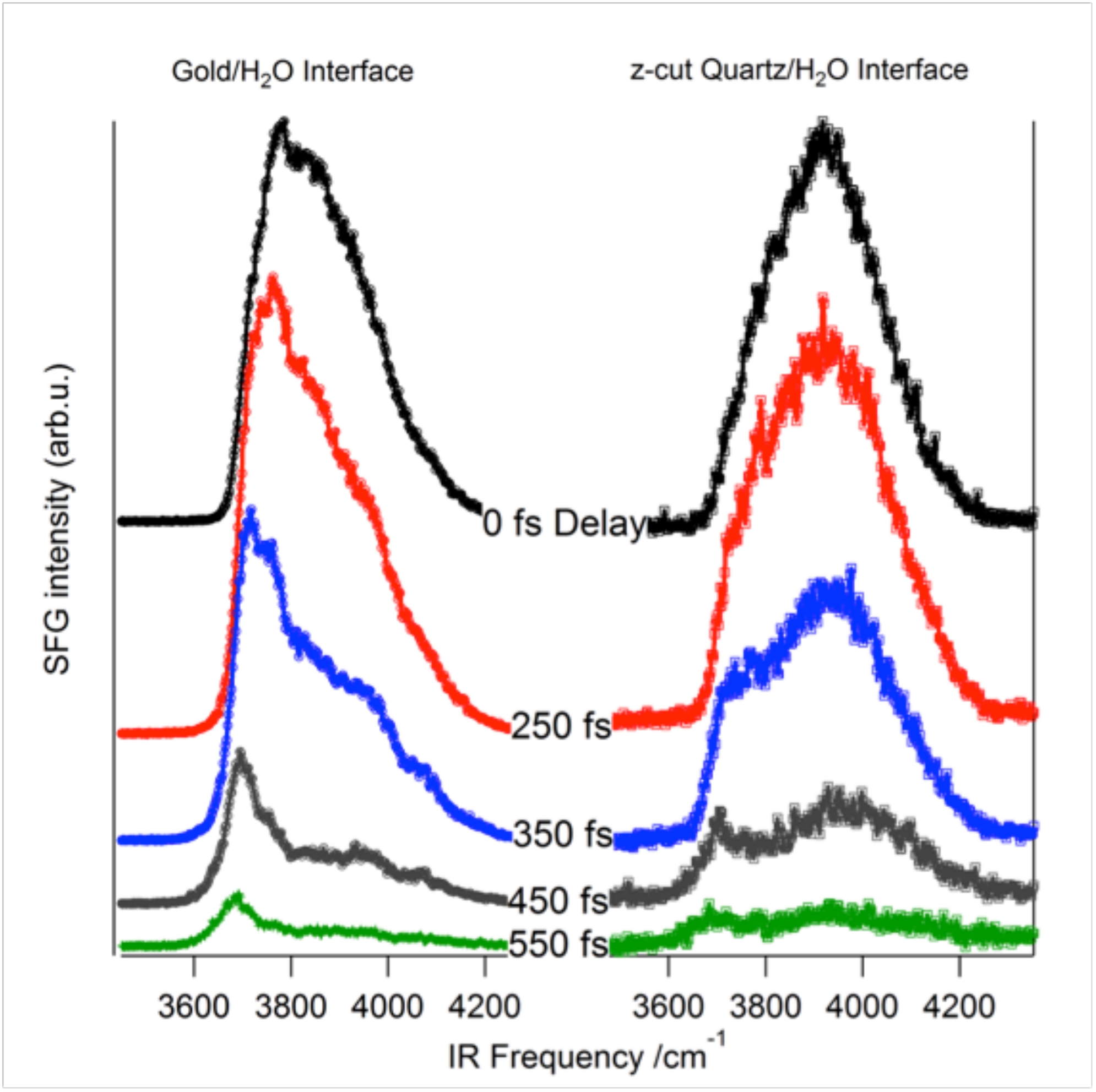}
	\caption{Delay dependent VSF spectra of Gold/water and z-cut quartz/water interface collected in the same sample cell. The resonant feature is present in at the Au, but not the z-cut qtz interface}\label{delay}
\end{figure}

As noted in the text, while effective at suppressing non-resonant contributions to the VSF signal the approach of Lagutchev et al \cite{lag07} -- delaying the visible upconversion pulse with respect to the infrared -- has been shown to introduce system dependent spectral distortions \cite{sti10}. Such distortions may result from chirp of the broadband IR as it passes through the water thin film in our spectroelectrochemical cell. We tested whether this chirp was responsible for the narrow peak we observed at the electrode/water interface by also collecting a non-resonant VSF signal from z-cut quartz under water as a function of delay between the IR and VIS pulses. Because our gold electrodes were deposited on z-cut qtz these measurements can be made by slightly shifting the sample cell on the laser table: no change in any optical components is necessary. As shown in Figure \ref{delay} the measurement from qtz as a function of IR-VIS delay does not show the strong feature apparent on gold. This comparison strongly suggests that the narrow feature we observe, after IR-VIS delay, at the Au/water interface does not result from chirp of the IR pulses as they pass through water. 

If this narrow feature at 3680 cm$^{\text{-1}}$ is a resonant signal from interfacial water we would expect that if we were to substitute D$_{\text{2}}$O for H$_{\text{2}}$O it should disappear and a feature at $\approx$2720  cm$^{-1}$ emerge. As shown in the inset in Figure 1(b) in the main text this is what we observe. It is in principle possible that the narrow feature we observe could also originate from water at the upper window of our thin-film spectroelectrochemical cell. Because there is no electrode on this window, if our narrow spectral feature reported on water at this interface we would not expect it to be potential dependent. As is clearly shown in Figure 2 in the text and in Figure \ref{pot_dep_SI} below, the spectral feature we observe after delaying is both strongly potential dependent and has a different potential dependence than the nonresonant signal. Note finally that the reflectivity of the gold electrode is somewhat potential dependent. As we show in Figure \ref{diode} below, this change is both too small and has the wrong sign to explain our data. Given these control experiments it seems clear that we have observed a resonant response of water at the gold electrode/aqueous interface.

\subsection*{Potential Dependent VSF Spectra}
\begin{figure}
	\includegraphics[width=0.9\textwidth]{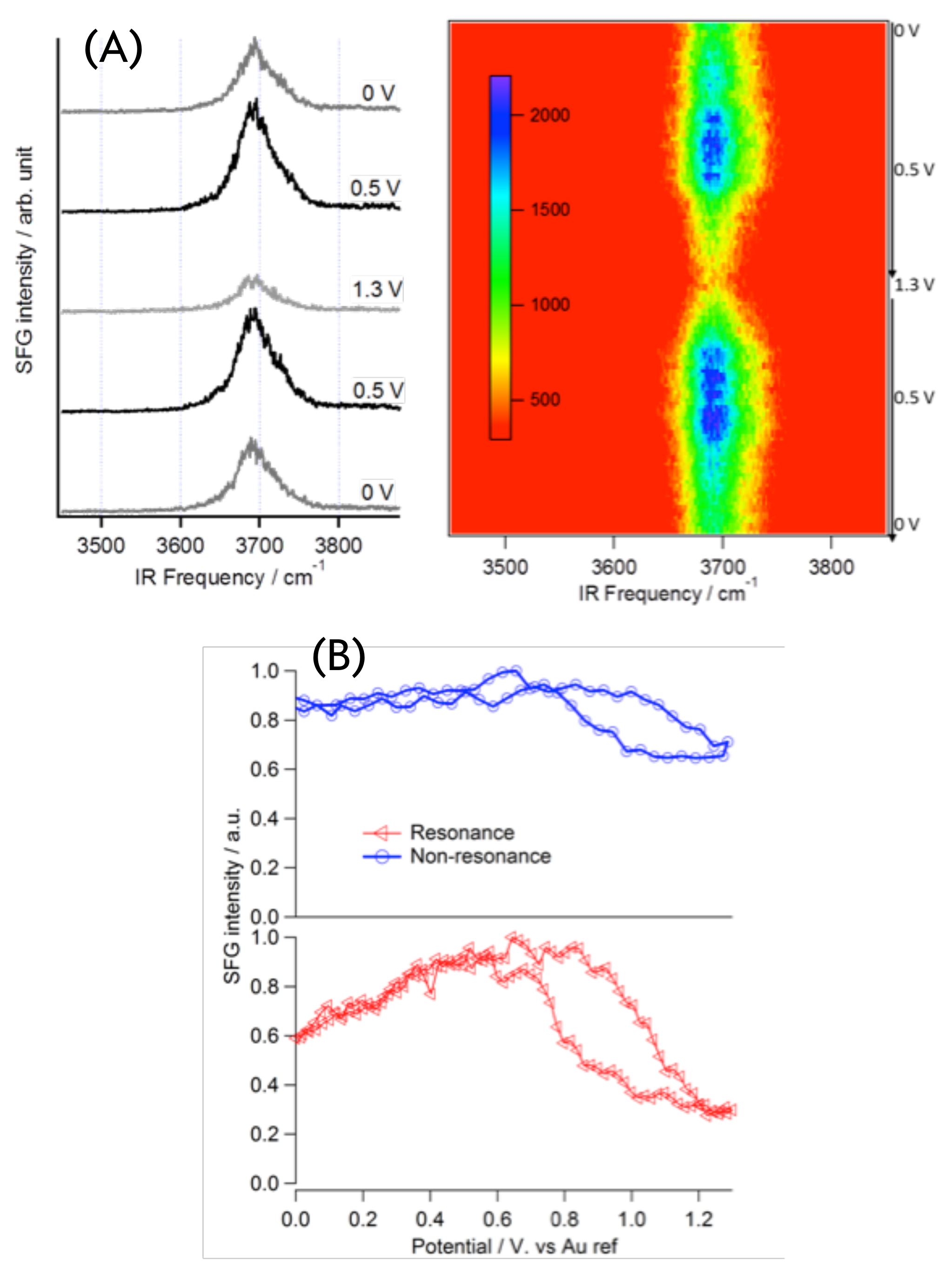}
	\caption{(A) Potential dependent VSF spectra showing a contour plot as a function of potential and selected spectra from particular potentials. Clearly only the amplitude changes significantly as a function of potential. (B) Integrated non-resonant and resonant VSF signal. Clearly the resonant signal exhibits a larger change with Au oxidation than does the non-resonant.}\label{pot_dep_SI}
\end{figure}

The corresponding potential dependent VSF spectra of the integrated intensity in Figure 2 of the main text are shown in Figure \ref{pot_dep_SI} (the delay between the IR and VIS pulses is 667 fs). It is clear from inspection that while the intensity changes dramatically with the applied potential, the center frequency of the peak is essentially potential independent: there is essentially no Stark shift consistent with a weak interaction between this OH group and the Au surface. Comparison of spectra collected at potentials near and further from the pzc does suggest that the spectral feature grows slightly more asymmetric (with a small shoulder on the high frequency side) as the surface loses charge. We envision two possible scenarios that may explain this feature of the data: near the pzc there is an additional, weakly interacting, type of OH group present at the interface that disappears as the surface is charged or, despite our delaying of the VIS and IR minimizing gold's nonresonant contribution to our measured I$_{\text{vsf}}$, there is a nonresonant contribution to our observable whose phase is potential dependent. Current work in our group focusses on the possibility of distinguishing these scenarios by heterodyning our interfacial water signal.

\subsection*{Potential Dependent Reflectivity of a Diode Laser}
The reflectivity of gold at visible wavelengths is potential dependent. To eliminate the possibility that this potential dependence explains the potential dependence of our resonant VSF signal we measured the potential dependent reflectivity of a laser-diode with a wave length of 614 nm (similar to our measured VSF wavelength). We find, as shown in Figure \ref{pot_dep_SI} that the reflectivity signal changes by 1\% over the range of interest with a qualitatively different trend than our observed VSF spectral response (see Figure 2(b) in the main text). 
\begin{figure}
	\includegraphics[width=0.5\textwidth]{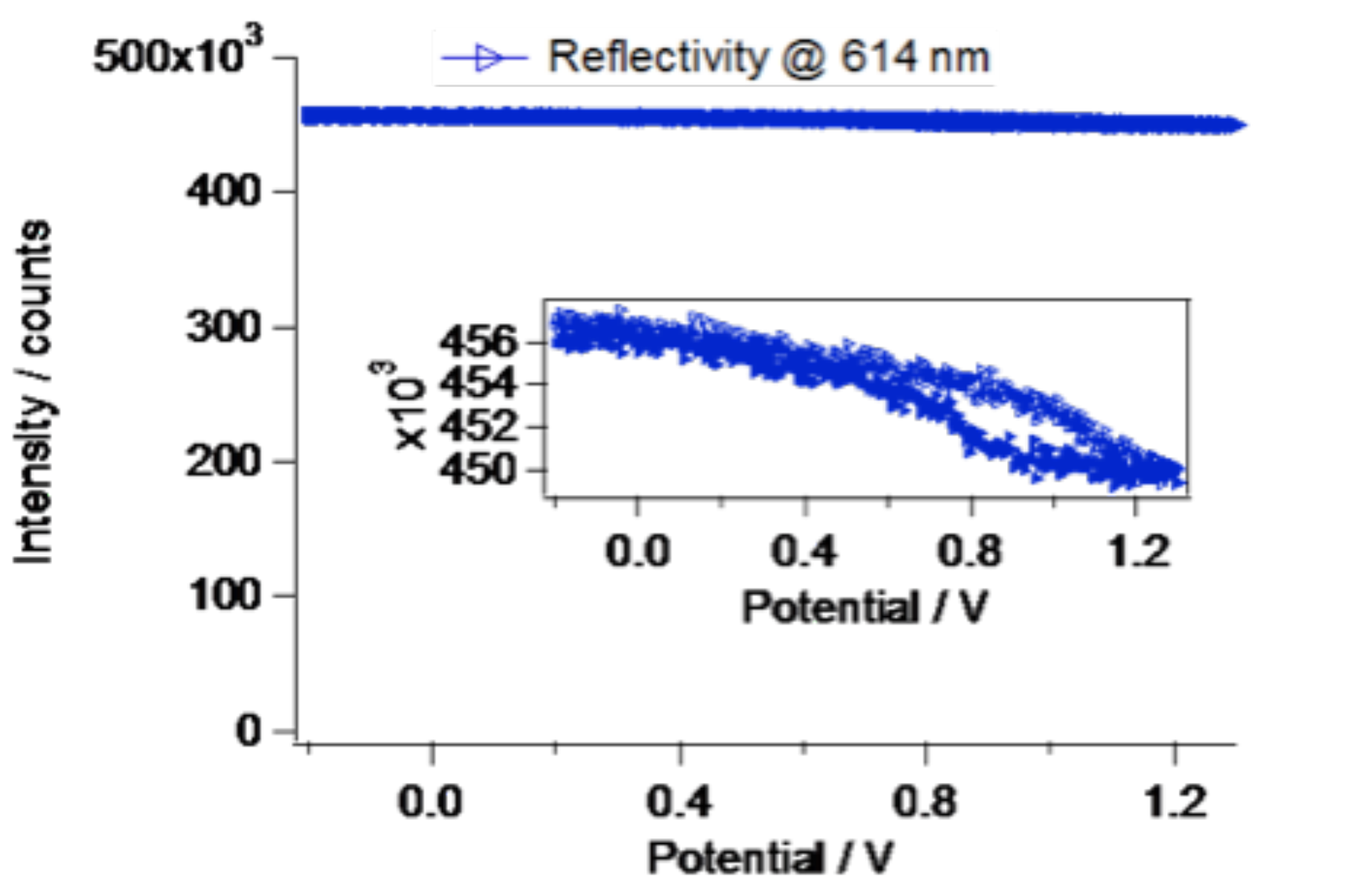}
	\caption{Potential dependent reflectivity change of the diode signal from the electrode surface.}\label{diode}
\end{figure}

\end{document}